\documentstyle[aps,preprint]{revtex}

\begin{document}
\title{Simplification of protein representation from the
contact potentials between residues}
\author{Jun Wang and Wei Wang}
\address{
National Laboratory of Solid State Microstructure 
 and Department of Physics, 
Nanjing University, Nanjing 210093, China}
\date{October 18, 2000}
\maketitle
\draft
\begin{abstract}
Based on the concept of energy landscape a picture of the 
mismatch between the reduced interaction matrix of residues
and the matrix of statistical contact potentials is presented. 
For the Miyazawa and Jernigan (MJ) matrix, rational groupings 
of 20 kinds of residues with minimal mismatches 
under the consideration of local minima and statistics on correlation
between the residues are studied. A hierarchical tree of groupings
relating to different numbers of groups $N$ is obtained, and 
a plateau around $N=8\sim
10$ is found, which may represent the basic degree of freedom of
the sequence complexity of proteins.
\end{abstract}



Using a small set of amino acid residues to reduce the sequence 
complexity in proteins has been made theoretically \cite{model} 
and experimentally\cite{R20,5lett}. Some patterns of
residues were discovered in the reconstruction
of secondary structures, such as binary patterns in
$\alpha$-helices and helix bundles \cite{R20} (see review
\cite{sim_pro} and references therein). These experiments imply
that not only the hydrophobic cores and the native structures but
also the rapid folding behaviors of proteins can all be realized
by simplified alphabets of the residues. These findings suggest
the existence of some small sets of residues for characterizing
the diversity of protein sequences. Theoretically, the simplest
reduction, the so-called HP model including H group with 
hydrophobic residues and P group with polar residues, has
been extensively used. Yet, the relation between different forms
or levels of these reductions (such as the $5$-letter palette
\cite{5lett}, or different forms of HP groupings
\cite{eigen_li,TD}) and the original sequences are not generally
established. To find out its physical origin is of importance for
the reduction of protein representation.

Previously, based on the Miyazawa and Jernigan (MJ)
matrix of contact potentials of residues \cite{MJ}, we made
reductions by grouping residues into different groups
\cite{wwnsb}. We found possible simplified schemes from
minimized mismatches between reduced interaction matrix
and the original MJ one. Here we report a physical picture 
of mismatch
based on the concept of energy landscape and some rational
groupings. Statistics on correlation
between the residues shows that some residues tend to aggregate
together or are friends to live in a same group. These enable us
to settle the groupings. 
A plateau of mismatch around group number $N=8\sim
10$ for three different interaction matrices is found, implying 
that groupings with
$N=8\sim 10$ may provide a rational set for protein reduction. This
coincides with a fact that proteins generally include more than
$7$ types of residues \cite{sim_pro}.

To divide $20$ types of residues into a number of
groups, the basic principle may be that the residues in a group
should be similar in their physical aspects, mainly the
interactions. After grouping, the residues in a group could be
represented by one of residues belonging to the group, thus the
complexity of protein sequence is reduced. When a residue is
replaced by another, the energy landscape of a protein should not
change its main feature (the shape) or the folding features are
basically the same. This is the case, especially when the system
is near the bottom of the funnel where a protein has the most
compact conformations. The energy difference
between two nearby conformations (c1) and (c2) is defined as
$\Delta E=\sum_n[ e_n^{(c1)}(s_i,s_j)-e_n^{(c2)}(s_k,s_l)]$ where
$e_n$ is contact energy of contact $n$ between two residues, 
$s_i$ is the residue 
type of $i$-th element in the protein sequence, and the number 
of contacts in two conformations are assumed to be the same. 
To keep the main feature of the energy landscape means that $\Delta
E$ should not change its sign, i.e.,
\begin{equation}
sign[\Delta E^{new}]=sign[\Delta E^{old}]
\end{equation}
when a residue $s_g$ $(g=i,j,k$ or $l)$ in $\Delta E$ is
substituted by one of its `friends' $s'_{g}$ in the same group.
Any discrepancy of Eq.(1) may change the energy landscape, and a
quantity ``mismatch'' is introduced to characterize the
discrepancy between the original protein and its substitute. Thus,
the mismatch acts as a quantitative non-fitness of substitutions
of residues for a certain grouping.

In details, $20$ natural residues are partitioned into $N$ groups
as ${G_1,\cdots ,G_N}$ (i.e., groups A, B and so on in
Ref.\cite{wwnsb}) with $n_i$ residues in group $G_i$, where
$n_1+n_2+\cdots +n_N=20$. Different values of ${n_i}$ give
different ``sets'' of the partition, and different arrangements of
residues into a given set represent different ``distributions'' of
the residues. The groups for a certain group number $N$
are represented as ${\cal G}_N =\{ \{ G^{(l)}_{K}(N), K=1,N\},
l=1,L_{N}\}$ where $G^{(l)}_{K}(N)$ means the $K$-th group in the
$l$-th set among the total sets with $L_{N}$ \cite{wwnsb}. For a
certain set, the mismatch will be minimized if the
residues are friends belonging to $N$ groups. [The residues which
are not aggregated together finally in a group are not friends.]
Due to the arbitrariness of contact index and various possible
distributions of residues, we define a strong requirement for a
successful grouping: no change of the sign for a substitution in
$\Delta E$, i.e., $\lambda (s_{i}s_{j}s_{k}s_{l})\equiv
sign[e(s_i,s_j)-e(s_k,s_l)]$ equals to $\lambda
(s'_{i}s_{j}s_{k}s_{l})\equiv sign[e(s'_i,s_j)-e(s_k,s_l)]$, e.g.,
when $s_i$ is substituted by one of its friends $s'_i$. Here
$s_i$, $s_j$, $s_k$ or $s_l$ belong to group $G_\alpha$,
$G_\beta$, $G_\gamma$ or $G_\delta$ with
${\alpha,\beta,\gamma,\delta}\in {1,2,\cdots,N}$, respectively.
Generally, when a residue is substituted by another residue
(friend or non-friend) from a same group well done or not well
done, one always has $\lambda (s'_{i}s_{j}s_{k}s_{l})=1$ or $0$ or
$-1$. Then, all possible substitutions give a sum of related
values of $\lambda$, i.e., $\Lambda (G_\alpha ,G_\beta
,G_\gamma,G_\delta )= \sum_{i}^{n_{\alpha}}\sum_{j}^{n_{\beta}}
\sum_{k}^{n_{\gamma}}\sum_{l}^{n_{\delta}}\lambda
(s_{i}s_{j}s_{k}s_{l})$ which describes the total effects of
substitutions of the residues from four groups of $G_\alpha$
,$G_\beta$ ,$G_\gamma$, and $G_\delta$. If
$\lambda(s'_{i}s_{j}s_{k}s_{l})$ is not the same as
$sign[\Lambda]$ (obviously $sign[\Lambda]=P$ in
Ref.\cite{wwnsb}), the substitution $s_{i}\rightarrow s'_{i}$ is
not favorable or the grouping of residues of $s_{i}$ and $s'_{i}$
in a group is a mismatch one. The average over all residues gives
out the total mismatch $M_{ab}$ of this distribution. Detailed
form of the mismatch see Ref.\cite{wwnsb}.

When the
element number $n_{i}$ in each group is fixed, different
distributions of residues in different groups may result in
fluctuant mismatches. Among all the distributions, the
best distribution (or the best arrangement of the residues) makes
a minimal mismatch $M_{abmin}$ for a certain set
$(n_{1},n_2,\cdots,n_N)$. To find out $M_{abmin}$, a Monte Carlo
(MC) minimization procedure \cite{wwnsb} is used. 
An enumeration over all possible distributions of residues can also
be made for small $N$.
With a fixed group number
$N$, we have a number of different sets which give different
minimal mismatches $M_{abmin}$. In principle, for a certain group
number $N$, we could chose the lowest mismatch and obtain the
related grouping as the final result among all sets $L_{N}$.
However this is difficult 
for those sets with many
groups with a single-element (MGWSE) or groups with singlets. For
example, as shown in Fig.1 for the set $(1,19)$ the mismatch is
the lowest among all $10$ sets (also the set $(1,1,1,1,16)$ for
$N=5$, and so on, see Fig.5). Obviously, this kind of mismatches
does not relate to the best or rational grouping of the residues.
Therefore, we must consider a local minimum (or plateaus) among
all sets as the rational global minimum $M_{g}$ (see Fig.1). 
Such a ``locality'' is motivated from the similarity
between two groupings. Two groupings are regarded as a couple of
neighbors when they can transform to each other just by exchanging two 
residues between two groups or by moving one
residue from one group to another. With this, all local minima (or
plateaus) are identified and analyzed. As shown in Fig.1, obviously 
there is a local minimum (or a plateau) besides those with MGWSE. 
Generally, different minima have different grouping patterns as
indicated in Fig.1. These local minima and plateaus may represent
some better groupings, and may reflect some intrinsic affinity
between the residues. As a result, they are taken as the
corresponding rational groupings with mismatch $M_{g}$. It is
worthy to note that for the grouping under some restrictions, such
as keeping the hydrophobic group unchanged \cite{wwnsb}, the
picture of the minimal groupings is the same although the grouping
space is limited.

The aggregation of some friendly
residues into a group  implies some correlation
between these residues. 
First let
us consider the two-residue correlation $C(S_i,S_j)$ by counting
the number of groups which include two residues $S_i$ and $S_j$.
That is, the count is taken as a quantitative scale of the
affinity between two residues, or a probability of two residues
being in a same group, among all groups in ${\cal G}_N$ for a
certain $N$, i.e., the groups for all sets $L_N$ with their
related $M_{abmin}$'s. Here the count $C$ is defined as
\begin{equation}
C(S_i,S_j)=\sum_{K=1}^{N}\sum_{l=1}^{L_{N}} I(S_i,G^{(l)}_K (N))
\times I(S_j,G_K^{(l)}(N))
\end{equation}
where $I(S,G)=1$ when $S \in G$, or zero when $S \notin G$.
Clearly, a matrix of the correlation between all possible pairs of
residues $C(S_i,S_j)$ can be obtained (see Fig.2). It is found
that the counts for some pairs are much large than those for other
pairs. This means that some residues are friends and some are
repulsion between each other, 
reflecting effective ``attraction'' between the residues in a group
and ``repulsion'' between residues in different groups. Note that
for the groupings of different $N$,
we have similar patterns.
The probability for finding a
certain group $G$ with specified residues among all the minimal
groups ${\cal G}_N$ can also be obtained by a count
$C'(G)=\sum_{K=1}^{N}\sum_l^{L_{N}}
D(G,G_K^{(l)}(N)) $
where $D(G_1,G_2)=1$ when $G_1=G_2$, or $0$ for $G_1 \neq G_2$.
As expected, different
groups have different chances to appear (see Fig.3). These
differences result from not only the grouping affinity between
residues but also the preference to the groups with a certain
size. For comparison, the count $C'(G)$ is normalized by the
total number of groups with the same size of group $G$ in the
statistical set ${\cal G}_N$. This normalized count is noted as a
probability of the occurrence for group $G$
\begin{equation}
P(G)=C'(G)/[\sum_{K=1}^{N}
\sum_l^{L_{N}} \delta(size(G),size(G_K^{(l)}(N)))]\, ,
\end{equation}
where $size(G)$ being the number of residues in group $G$, and
$\delta (size1,size2)$ being the $\delta$-function. From Eq.(3) it
is found that some groups have large probabilities $P(G)$ and
appear many times with large number of the counts $C'(G)$, implying
that the residues in these groups having more chances to be in a
group or that these groups having strong preference to appear in
the grouping. Thus the grouping with these groups shows a better
settlement of 20 kinds of residues than others. Note that there
are some groups with high probabilities $P(G)$, but small
statistical counts $C'(G)$. Such groups generally have large
numbers of elements and only appear one or two times in 
${\cal G}_N$, which makes the normalization factor in Eq.(3) rather small.
Clearly, these groups are removed in our analysis because of
lacking of the statistical reliability.

For the MJ matrix, 
as
shown in Fig.4, the groupings follow a hierarchically tree-like
structure. That is, $20$ kinds of residues are
firstly divided into two groups (also see Fig.1a), namely the H
group with residues ({\it C, M, F, I, L, V, W, Y}) and the P group
with residues ({\it A, G, T, S, N, Q, D, E, H, R, K, P}). Then the
H and P groups are alternatively broken into two or more groups
relating to different $N$. For example, for the case $N=3$, the P
group are divided into two small groups, i.e., ({\it A, H, T}) and 
({\it G, S, N, Q, D, E, R, K, P}). 
For the case $N=5$, the H group is divided into ({\it F,
I, L}) and ({\it C, M, V, W, Y}), and the P group is divided into
({\it A, H, T }), ({\it D, E, K}) and ({\it G, S, N, Q, R, P}),
respectively. Similar results are obtained for $N$ up to $9$ with
a sequential order of hydrophobicity without any overlap
between the hydrophobic branch and the hydrophilic one following
the H/P dividing. The difference between the
present study and the previous one in Ref.\cite{wwnsb} is that
there are alternant breaking of the H and P groups in the new
groupings, which gives out a little decreasing in the mismatches,
and also slight different representative residues.

Our new analysis relates to a clearer physical
picture of the rational groupings. Following the
tree-like groupings, one can see the dividing on the H
groups or the P groups (see Fig.4). For the case of $N=3$, to
divide the P group (on the base of $N=2$) is
obviously more rational than to divide the H group, suggesting a
priority for dividing the P group first.
Differently, for the case of $N=4$, we should divide the H group
first, and then for the case of $N=5$ divide the P group again. It
is found that the dividing is alternant, reflecting the detailed 
differences between the interactions of the H
and P groups.
The former results under some restrictions, such as to
fix the H group (with $8$ residues) unchanged, may relate to
somewhat rough dividings, resulting in large mismatches (see the
data for $N=3$, $4$, and $5$ in Ref.\cite{wwnsb} and Fig.5). 

 Fig.5 shows a monotonic decrease in the mismatch for
$N=2$ to $20$, which implies the more groups the better. Besides,
there is a plateau near $N=8$ in this curve (case-A),
which characterizes the saturation of the grouping. 
This means that more groups will not further
decrease the mismatch or more groups might not greatly enhance the
efficiency of the complexity reduction. Thus the number $N=8$ may
indicate the minimal number of types of residues to reconstruct
the natural proteins, or a basic degree of freedom of the
complexity for protein representation. This, in some sense,
relates well to the result in the previous studies \cite{wwnsb},
and an argument in Ref.\cite{sim_pro}. Noted that the former
plateau at $N=5$ ceases due to the canceling of the grouping
restriction.
Interestingly, in Fig.5, we also plot all the lowest mismatches
relating to the groupings with MGWSE which generally are not the
local minima as discussed above. A typical example is the grouping
with groups $(1, 1, 1, 1, 16)$ with a mismatch $M=0.04747$, which
is the lowest one among all sets of $N=5$. However, it is noted
that even including all these trivial cases for $N=2$ to $14$, the
curve still shows a plateau around $N=9$ with eight groups with
single residue of {\it C, M, F, I, L, V, W, Y} and one group with
the rest residues as well. Clearly, this
plateau relates again to the saturation of the H and P grouping or
the detailed differences between the residues of the interactions,
and also gives out a support on the discussion for the $N=8$
plateau above. As shown in Fig.5(b), we have similar results
for two other interaction matries \cite{TD,SW}.

Finally, we note that for each grouping with different $N$,
we have found the representative residues for the MJ matrix,
e.g., ({\it I, A, D}) for $N=3$, ({\it I, A, C, D}) for $N=4$
and ({\it I, A, G, E, C}) for $N=5$. The slight change in the 
representative residue for $N=5$ is attributed to the different
implementation and requirement in the grouping. The foldability
and the effectiveness have also been studied, we will report on these
elsewhere.

In conclusion,
we present a grouping method based on the
requirement that energy landscape is basically kept in the
reduction. A quantity, the mismatch, is taken as the measurement
of the goodness for the reduction.  
Our results imply that the residues
do have some similarities in their interaction properties and can
be put together into groups. Then by choosing a single residue for
each group, the complexity of proteins can be reduced or the
proteins can be represented with reduced compositions. 
Especially, a basic degree of freedom of the complexity with $8\sim 10$ 
types of residues is found.

This work was supported by the Foundation of NNSF (No.19625409, 10074030).
We thank C. Tang, C.H. Lee and H. S. Chan for comments and suggestions.

\begin{figure}
\caption{
The mismatch $M_{abmin}$ for different
sets for $N=2$ (a) and $N=3$ (b). The set index 
represents the sets marked in the figure.
} 
\end{figure}

\begin{figure}
\caption{
A two-residue correlation statistics for
Eq.(2) for the MJ matrix. Different shades of gray represent
different values of the count $C(S_{i},S_{j})$ among all 527
groups for $N=5$. 
}
\end{figure}

\begin{figure}
\caption{
Probabilities $P(G)$, i.e., Eq.(3), and
the counts $C'(G)$ of occurrence of groups $G$ with
$N=5$. The group index is arranged following the
magnitude of the probability of the groups. Some groups are
labeled.
}
\end{figure}

\begin{figure}
\caption{
The rational groupings of a
hierarchically tree-like structure for the MJ matrix for $N$ 
up to $9$.
}
\end{figure}

\begin{figure}
\caption{
The minimal mismatch $M_{g}$ vs. $N$: 
(a) for the MJ matrix;
(b) for contact potentials in Ref.[6] (TD case)
and in Ref.[9] (SW case). The plateaus are shown for
different cases.
}
\end{figure}

\end{document}